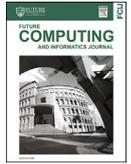

# Privacy-preserving data aggregation in resource-constrained sensor nodes in Internet of Things: A review

Inayat Ali*, Eraj Khan, Sonia Sabir

*COMSATS Institute of Information Technology, Abbottabad 22060, Pakistan*

## Abstract

Privacy problems are lethal and getting more attention than any other issue with the notion of the Internet of Things (IoT). Since IoT has many application areas including smart home, smart grids, smart healthcare system, smart and intelligent transportation and many more. Most of these applications are fueled by the resource-constrained sensor network, such as Smart healthcare system is powered by Wireless Body Area Network (WBAN) and Smart home and weather monitoring systems are fueled by Wireless Sensor Networks (WSN). In the mentioned application areas sensor node life is a very important aspect of these technologies as it explicitly effects the network life and performance. Data aggregation techniques are used to increase sensor node life by decreasing communication overhead. However, when the data is aggregated at intermediate nodes to reduce communication overhead, data privacy problems becomes more vulnerable. Different Privacy-Preserving Data Aggregation (PPDA) techniques have been proposed to ensure data privacy during data aggregation in resource-constrained sensor nodes. We provide a review and comparative analysis of the state of the art PPDA techniques in this paper. The comparative analysis is based on Computation Cost, Communication overhead, Privacy Level, resistance against malicious aggregator, sensor node life and energy consumption by the sensor node. We have studied the most recent techniques and provide in-depth analysis of the minute steps involved in these techniques. To the best of our knowledge, this survey is the most recent and comprehensive study of PPDA techniques.
Copyright © 2017 Faculty of Computers and Information Technology, Future University in Egypt. Production and hosting by Elsevier B.V. This is an open access article under the CC BY-NC-ND license (http://creativecommons.org/licenses/by-nc-nd/4.0/).

## 1. Introduction

Ubiquitous sensing has already been evident due to the proliferation of sensor devices in Wireless Sensor Network (WSN) and pervasive availability of internet in the form of Wi-Fi and mobile cellular internet (3G, 4G). This ubiquitous sensing of devices and their interconnection in communicating and actuating manner creates the Internet of Things [1]. The notion of the internet of things is the interconnection and co-ordination of everyday smart internet-enabled objects such as sensors, smartphones, actuators, smart watches, etc. — to each other's and to humans via some mean of connectivity to reach common goals [2,22]. IoT is fueled by existing technologies of Wireless Sensor Network and Radio Frequency Identification (RFID). IoT covers the application areas of WSN and RFID with an addition of accessibility through the internet. IoT has many application areas such as smart healthcare system, smart home, smart retail industry, smart logistics, smart transportation system and the list goes on [1]. Though IoT is a breakthrough innovative idea that will bring ease and comfort to human life, still it faces many challenges in its way to being widely accepted by the users. The major challenge to IoT is security and privacy of the user's [7]. For the sake of argument, we discuss privacy problem in two application areas of IoT i.e. Smart Healthcare system (WBAN) and Smart Grid.

Smart healthcare system or e-health is an application domain of IoT that is powered by the technological

\* Corresponding author.

*E-mail address:* falcon19khan@gmail.com (I. Ali).

Peer review under responsibility of Faculty of Computers and Information Technology, Future University in Egypt.

convergence between Wireless Body Area Network (WBAN), WSN, Mobile Crowd Sensing (MCS) and cloud computing [3] as shown in Fig. 2. In e-health, the patient's health is monitored continuously using sensors attached to the body of the patients. The health of the patient is monitored in real time, whether he/she is in the hospital for some routine checkup, in home, office or even if the patient is walking in a market. The e-health is very favorable application domain of IoT that will revive the traditional healthcare system. However, the health-related data of a person is very confidential and also critical in some cases. Any leakage of such confidential data leads to serious privacy issues. A patient medical record is collected in real time through the WBAN of the patient and communicated to the cloud for storage via the internet. WBAN consist of resource-constrained sensor nodes. The data is accessed by the authorized medical personals and a report for proper medication can be generated and pushed to the patient smartphone. However, the medical record of the patient can be used for malicious purposes by the adversary. Therefore, the health-related data must be kept confidential when in transit, use and when in storage. Many advanced countries have legislation to protect the privacy of health-related data. HIPAA (Health Insurance Portability and Accountability Act of 1996) [4] is US legislation that provides data privacy for health-related data, which reflect the importance of the health data privacy. Nonetheless, this data must also be technically protected from adversaries when data is in transit between the constrained sensor node and its gateway. Similarly to increase sensor node life data aggregation is used to reduce communication overhead and upsurge sensor life. However, the data aggregation creates serious privacy problems of data from the aggregator and hence the data privacy must also ensure from the aggregator nodes during aggregation.

Smart grid, another prominent application area of IoT is still not widely accepted by the users due to privacy concerns. In the Smart grid, the energy measurement of a building is measured continuously and sent to the utility provider for billing and other power management purposes to cope with the high demand for electricity. Smart meters are an important entity in the smart grid that collects power consumption readings from electric appliances in the home or any other commercial building and sends the aggregated result to the utility provider for billing and management purposes [13]. However, the data is very sensitive and can be used for user profiling if leaked or available to the adversaries. The adversary could be able to know about the users' private life and habits and can also harm them using this information [13]. Different techniques have been proposed to ensure the privacy of data in smart grid [10,11,13], yet further efforts are required to design techniques that overcome the limitations of existing mechanisms.

The sensor nodes in WBAN, WSN, and smart grid are often resource constrained and battery operated. Power hungry cryptographic operations for security and privacy and communication overhead shrink sensor life. Data aggregation is the solution to reduce communication overhead and energy consumption at each sensor node, but it suffers from the problem of data privacy when the data is aggregated at each node or the aggregator nodes [5]. Efficient data aggregation can surge the life of sensor nodes and the sensor network, as it reduces computation at each node and communication overhead in the network Fig. 1. Therefore, to take benefit from data aggregation to increase sensor network life, different privacy-preserving data aggregation techniques have been proposed [5,8–13,16–21]. Privacy-preserving data aggregation techniques reduce the energy consumption at each node by lowering the communication overhead while preserving the privacy of sensor data at the same time. Different techniques in this area have been proposed so far, but each one has its own limitation as discussed in section 2.

Application areas of IoT that are based on resource-constrained sensor nodes need to gain users' trust to be accepted by them. Smart grid technology has been rejected by the British government once due to privacy problems. Smart healthcare also exists just in theory and is not in practice due to the same privacy concerns. Many efforts had been made so far by the researcher to overcome the problems of privacy. In this paper, we discussed the latest PPDA techniques and compare their performance as to guide the researcher to where efforts should be made to develop PPDA techniques for resource-constrained IoT. To the best of our knowledge, this is the latest survey to discuss these PPDA solutions. In addition, this review provides very in-depth and critical analysis of each mathematical operation involve in different PPDA schemes. This critical analysis and comparison will help the researchers to design energy efficient and computationally feasible solution to ensure user's privacy in IoT applications.

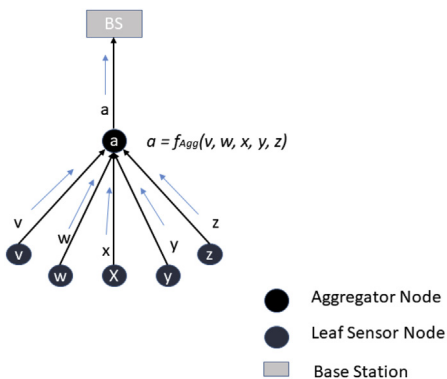

Fig. 1. Data aggregation in sensor network.

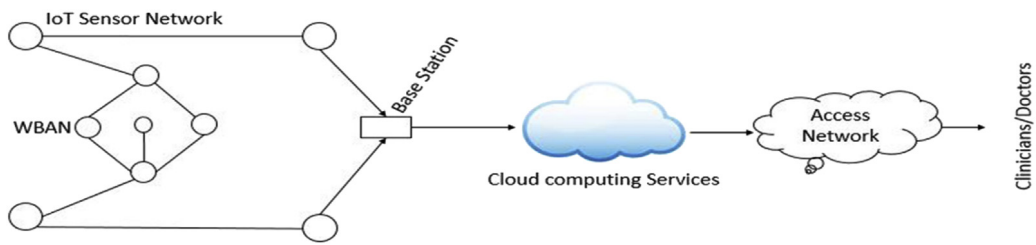

Fig. 2. e-Health monitoring system.

The remaining paper is structured as follow, section 2 discuss the various state of the art PPDA techniques in resource-constrained sensor nodes. In section 3 we provide a comparative analysis of the techniques and finally, we conclude the paper in section 4.

## 2. Privacy-preserving data aggregation techniques

There are many PPDA techniques proposed for WBAN, WSN and smart grid. We selected the latest PPDA techniques to discuss and compare their performance. The techniques are discussed below and summarized in (Table 1).

Bista et al. [9] proposed a scheme to preserve the privacy of WSN data from internal security threats. The existing schemes CPDA and SMART [12] have high communication cost, while the proposed scheme is efficient in terms of communication overhead and power dissipation.

The scheme uses the additive functionality of complex numbers to aggregate data and masks the sensitive data by using a random number that is a shared secret between Query server (sink) and each sensor node. Each node after sensing data, add it to its random number to mask sensitive data, change it to complex number format by adding it with a complex number and then encrypt it with a symmetric key before sending it to another intermediate node (aggregator). The intermediate node decrypts the data, aggregate it by adding the data received from other nodes and encrypt it again with a symmetric key before forwarding it to another intermediate aggregator. The process goes on till the customized aggregated data reach to the query server (QS). The QS separate real and imaginary part before issuing the final result. The original sensor data lies in the real part of the aggregated data.

This scheme is compared with existing schemes CPDA and **SMART** [12] and it is shown that this scheme gives better performance in term of Energy Dissipation and communication overhead.

Though this scheme is better in communication overhead and power dissipation, the authors of this paper did not consider the communication overhead generated by Key exchanges and management between the nodes and the power dissipation by the encryption algorithms at each node, as each node perform both decryption and encryption for each message it received.

Yip et al. [10] use Incremental Hashing Function (IHF) to develop a scheme for Privacy-preserving and Cheat-Resilient (PPCR) electricity usage and reporting for Smart Grid (SG). The authors argue that IHF is suitable for resource-constrained smart meter as IHF require low storage and computation power.

Initially, the smart meter will calculate the cost per load consumed and hash the result with IHF before sending it to the Operation Center. Operation center upon receiving the hashed cost from all the meters will sum all the hashed cost and send it to the utility provider. The utility provider checks the integrity of the report by comparing the hashed value sent by operation center and the hashed value of total power generated for that neighborhood. At the end of the billable period say 24 h each consumer again send its total consumption report which is validated by the utility provider in the same way before billing the consumer.

The scheme provides privacy as if the hashed cost of a consumer in near real-time is eavesdropped by an adversary or even the operator at operation center, it is impossible to find the input cost due to the pre-image resistant property of IHF. The scheme also provides authentic reports to the utility provider as utility provider check the integrity and authenticity of the report by comparing the summed hashed report of a neighborhood sent by Operation center and the total hashed of power generated for that neighborhood. The scheme is efficient in term of computation complexity as its complexity is only O (N), and also require low fixed size storage as IHF generate a small fixed size hash value for any input size.

The scheme performs data aggregation and hashing at smart meter, so it gives high security and privacy at this level. However, it does not provide a solution if data is overheard at device layer (communication before the smart meter or from appliances to smart meter) by the adversary.

Finster and Baumgart [11] proposed a new scheme for privacy-preserving data aggregation, which is an extension of the existing techniques for Wireless Sensor Networks "SMART" [12]. However, the SMART techniques do not consider performance degradation due to communication errors. The author extended the SMART techniques for smart grid metering by adding the exactness and robustness in SMART in the presence of communication errors. The author, therefore, proposed SMART-ER for smart grid privacy.

SMART consist of three steps (a). Slicing: Each node slices its private data into N+1 slices and sent all the slices to other nodes keeping one with itself. (b). Mixing: Every node after receiving slices from other nodes aggregate them and send it to the data sink. (c). Aggregation: The sink node after receiving slices from each node sum them to get its required data. The data sink is not able to extract individual node information from summation as summation process is irreversible.

SMART-ER is robust against communication errors and node failures, unlike SMART. It makes the following changes in SMART to solve the deficiency of SMART.

1: Slicing using completely randomized data.
2: Dependency tracking
3: Grouping of nodes into smaller groups to stop error spreading.

4: Extrapolation to recover lost submission.

While randomizing the slices SMART-ER performs pre-slicing without knowing the private data. Pre-slicing also reduces the delay at the data sink.

For correctness, the scheme finds nodes dependencies and discard the value at sink whose dependency nodes did not send their submission or lost on the way. The algorithm also divides nodes into smaller groups to shortened dependency node set. At the end, the algorithm extrapolates the loss nodes data as sink known the total number of nodes. SMART-ER is better than SMART with regards to a number of valid submissions. SMART-ER give 85% of valid submission while SMART gives only 10% valid submission. Communication overhead in SMART-ER is a bit higher than SMART but still, it is negligible.

Though the scheme improved SMART performance for Smart Grid but it estimates the number of missing nodes information by extrapolation that is just an estimated value and maybe not exact most of the time. Because it is difficult to extrapolate a value with less amount of available nodes data.

Callegari et al. [13] proposed a distributed architecture for communication in the smart grid that is privacy aware. The architecture is able to anonymously process the energy production and consumption data and not revealing the user identity information. The architecture is based on secure multi-party computation (SMPC), verifiable secret sharing (VSS) scheme and makes use of efficient zero-knowledge (ZK) tool. The scheme is intended to secure high frequency data of smart meters that can be used for user profiling due to its sensitivity.

The proposed architecture consists of smart meters, privacy peers, and utility server. Smart meter collect high frequency sensitive data and low frequency data for billing purposes which are not sensitive and cannot be used for user profiling. Smart meter sends low frequency data directly to the utility servers and high frequency data to the privacy peers, Privacy peers located between smart meters and utility server. Both privacy peers and smart meters are considered to be not trusted entities while the utility servers are a trusted entity that is used for billing, account management, activities operation and user profiling.

At smart meter, a data vector "$v$" is generated by subtracting uniformly generated random vector "$u$" from original data vector "$b$" of the user. Vector "$v$" is then divided into "$n$" shares each for one peer and sent to them. Privacy peers do not have information about "$d$" of the user. Privacy peers after receiving shares from all smart meters calculate a composite share and send it to the utility provider.

The utility provider first sums all "$u$" received directly from smart meters. It also sums all the shares received from all privacy peer of a neighborhood. Summation of both values gives the composite shares of the whole neighborhood "$D$" by homomorphic properties of the secret sharing scheme. In this way, utility provider is able to construct the high frequency data of each neighborhood that can be used to construct individual user profile of the neighborhood.

The author claim that the architecture is secure against compromised privacy peers attack and cheating of user on smart meter.

If the privacy peers for a neighborhood are compromised and composite shares leaks. The adversary can easily eavesdrop on "$u$" shared by each user directly with the utility server over a non-secure channel. An adversary can get "$D$" by summing all the "$u$" and composite shares from privacy peers.

To enhance network nodes lifetime in WSN, different techniques are used. Among which data aggregation is an efficient technique to reduce communication overhead and increase network nodes lifetime. However, data aggregation face challenges of data integrity. The aggregator is unable to verify the validity of the packets received which is a vulnerability that can be exploited by the attacker to inject its malicious packets. To solve this problem homomorphic MAC techniques have been proposed by Hayouni et al. [5]. The schemes used two MACs for each encrypted message, one is MAC for data integrity and the second one is EMAC for MAC integrity. The scheme contains four phases, Key distribution, Encryption, MACs, EMACs construction, and combination. Two sets of keys containing R subkeys are generated and sent to each source node. Each message is then divided into m generations and encrypted. After that MACS and EMACs are constructed and then combined with a message to create a final message to forward. This scheme provides both privacy and integrity but uses too many keys during its operation at each source node. Key management is an issue of the proposed scheme.

Kumar and Madria [8] proposed a novel energy efficient algorithms that are used to preserve data privacy as well as data integrity in the course of data aggregation. The algorithm is based on Recursive Secret Sharing (RSS), where the shares of the data $\delta$ are used to store $k - 2$ additional pieces of information. A node with at least k shares can easily reconstruct all of the $k - 1$ pieces of hidden information. This algorithm provides a construction in which we can prevent a node which has all the shares, from reconstructing and retrieving the hidden data. Each node creates shares for its data using Recursive secret sharing (RSS) and sends all the data to its parent after scrambling the data with a key to protecting from curious aggregator. An integrity share keys between nodes and BS is used to verify the integrity of the data. This integrity key is embedded in the share using RSS. The scheme consists of two algorithms, one for generating the shares of the data and the keys and the second one is to regenerate data and integrity check at BS. The assumption of the scheme is that sensors form a tree like network structure. In Share generation algorithms sensor reading $\delta k$, PRF f (.), nonce n, random seeds r1, r2, r3, r4 are used to generate perturbation key, scrambling key and integrity key. These keys will be used instead of random seeds in the next iteration. Perturbed data is generated using $\delta \eta k = \delta k + \eta k$ where $\eta k$ is perturbation key. RSS is then used and a linear combination of the integrity key I? k + I?? k $\delta \eta k$ and data, $\delta \eta k$ are encoded to create the shares $\omega 2\ k(3), \omega 2\ k(4)$, and $\omega 2\ k(5)$. These shares are then scrambled with scrambling key to generate scramble shares $\lambda 1k, \lambda 2k$, and $\lambda 3K$ and send to the parent which aggregate them and send them to the BS. The BS the use the sum of the scramble keys to unscramble the aggregates shares. These shares are then used to decode received perturbed data.

The algorithms are very efficient in terms of energy consumption, memory usage, bandwidth consumption and execution time but still, it is based on the assumption of a tree like structure made by sensor nodes. It also does not discuss the mechanism to feed the variables (random seeds, nonce, and prime number) in sensor nodes.

To ensure data privacy during aggregation and enhance data transmission efficiency, Othman et al. [19] proposed a technique based on homomorphic symmetric encryption. The scheme also ensures data integrity using homomorphic signature.

Zhang et al. [16], proposed a privacy-preserving and priority based data aggregation and forwarding scheme for the mobile healthcare system. The main focus of the paper is to increase the delivery ratio of the health data to the cloud. Nevertheless, the paper also preserves the privacy of the data by using Paillier cryptography. Paillier cryptography is asymmetric cryptosystem based on public and private key which is comparatively computation extensive for tiny and resource constrained sensors.

Othman et al. [17] proposed data aggregation technique that is energy efficient, to ensure data confidentiality and integrity of data that is secure against node compromise attack. The author uses Elliptic Curve Okamoto-Uchiyama (EC-OU) for data confidentiality and Elliptic Curve Digital Signature Algorithm (ECDSA) for ensuring data integrity during data aggregation in WSN.

To reduce the performance degradation due to malicious data injection and its verification at the base station, Rafik and Mohammed [18] proposed novel techniques that verify the data at each hop thus providing early detection of attack and preserve end to end privacy. The authors used Elliptic Curve El Gamal (ECEG) cryptosystem for data confidentiality which is public key cryptosystem. They also use homomorphic property of ECEG to ensure confidentiality of data from aggregator node also.

## 3. Comparative analysis

In security algorithms for IoT motes, the most important performance parameters are computation cost, communication overhead, sensor life, Privacy or security level, privacy against malicious aggregator and total energy consumption at each constrained IoT sensor node. The importance of these parameters is owing to the resource-constrained sensor node design in IoT. The above mentioned parameters are the desired features of any security algorithm in the Internet of Things. In this literature, we will use the abbreviation CC, SL, CO, PL, PA and EC for computational complexity, sensor node life, communication overhead, privacy level, privacy against aggregator and energy consumption respectively. For the bar graph comparison below in Figs. 3—6, we have found the number of mathematical operations in each algorithm and estimated the computational complexity of each operation to assign weights to these operations. We have estimated the relative value of CC based on these assigned weights. Such as we assigned a weight of 5 for each PKC and 2.5 for symmetric key encryption and decryption and a constant 1 for each addition/subtraction operation.

Similarly, for CO we have estimated the communication overhead by assigning weights to different operations like data slicing generates a high CO and thus we assign a relative value of 6 and a constant 2 for the addition of a tag/hash to a message and so on. We also used a similar formula for estimating sensor node life, privacy level, and energy consumption.

### 3.1. Computational cost

In computational cost (CC) we will analyze the algorithms based on the complexity of mathematical steps involved. The techniques proposed by Hayouni et al. [5] has a very high CC because it involves three mathematical operations, first, each message is encrypted at source node using a key vector and mod operation. Mod operation includes repeated multiplication and reduction that has high CC. Second and third operation is to calculate MAC and EMAC. Combining all the operations makes the algorithms computationally very complex.

Yip et al. [10], Finster and Baumgart [11], Kumar and Madria [8] and callegari et al. [13] shows closed performance to each other's as all of theme use some form of secret sharing and data slicing. The CC of [10] is a bit high than [11] and [13] because [10] used Incremental Hashing Function (IHF) for the integrity of data while [13] and [11] divide the data into $n$ secrets for privacy. Kumar and Madria [8] use recursive secret sharing for generating share which involves polynomial expansion. Polynomial expansion has CC of O ($n$). Bista et al. [9] involve repetitive encryption and decryption at each node and thus it has medium CC owing to symmetric key encryption which are computationally light compare to Asymmetric cryptography.

The techniques proposed in Refs. [16—21] all are based on public key cryptography. Public key cryptography is computationally very expensive and that's why all of these techniques have very high computation cost. Among these techniques [16,18,21] use Public Key Cryptography (PKC) only for encryption of messages and thus has high computational cost while the remaining [17,19,20] use PKC for encrypting messages and also for digital signature and thus it possesses very high computation cost. The CC of each of these techniques can be viewed in Figs. 3—6 below.

### 3.2. Communication overhead

Communication Overhead (CO) in Hayouni et al. [5] is also high as it distributes two sets of key vectors to each node for a certain time and also adds two tags to each message for integrity, which create extra burden in the network. CO of [13] is a slightly high than [10] as it sends all the secret to privacy peers while [10] calculate the total cost and send it not the original data to the utility provider and thus it has a very lower CC and CO as shown in Figs. 3–6. However the scheme in Ref. [10] could not be used in most of the applications because it do not send the original data but rather a calculated cost value. Finster and Baumgart [11] have high CO because it divide the data into $n+1$ pieces and send them to $n$ nodes. Which means that a single message will now take the overhead of $n+1$ headers. Moreover each node also communicate dependency vector to BS at the end of each message. Thus CO in Ref. [11] is very high. CO of [8] is medium as compared to [11] because in Ref. [8] shares contain $k-2$ additional piece of information. Since techniques proposed by Bista et al. [9] use symmetric key encryption that means that $n(n-1)/2$ keys are required for $n$ users and these keys are to be updated after an interval that create extra overhead in the system.

Table 1
Summary of techniques.

| S.# | Paper title | Approach | Description | Limitation |
|---|---|---|---|---|
| 1 | A Novel Efficient Approach for Protecting Integrity of Data Aggregation in Wireless Sensor Networks [5] | Message Authentication Code | Use Two MACs for each encrypted message. One MAC for data integrity and second for MAC integrity, Encryption at each node | Uses too much keys, Key management |
| 2 | A Privacy-Preserving and Cheat-Resilient Electricity Consumption Reporting Scheme for Smart Grids [10] | Incremental Hashing Function | Perform data aggregation and hashing at smart meter, Cheat Resilient electricity Usage | No privacy if the data is overheard before smart meter |
| 3 | SMART-ER: peer-based privacy for smart metering [11] | Data Slicing | Robust communication in smart grid, reduce communication errors | Extrapolated value is hard to be correct because of less number of missing nodes data |
| 4 | Confidentiality and Integrity for Data Aggregation in WSN Using Homomorphic Encryption [19] | homomorphic Asymmetric encryption | Homomorphic encryption and signature for privacy and integrity | Elliptic curve cryptography (Asymmetric algorithms), Boneh signature scheme that can be aggregated (Public Key Cryptography) |
| 5 | PIP: Privacy and Integrity Preserving Data Aggregation in Wireless Sensor Networks [8] | Secret Sharing | Prevent nodes from reconstructing and retrieving the hidden data even having all the shares | Do not discuss the mechanism to feed the variables in sensor nodes. |
| 6 | PHDA: A priority based health data aggregation with privacy preservation for cloud assisted WBANs [16] | Asymmetric cryptography | Privacy preserving and priority based data forwarding, Improve the delivery ratio of data. | Asymmetric cryptosystem. Which is computational extensive and not suitable for tiny nodes |
| 7 | A distributed privacy-aware architecture for communication in smart grids [13] | secure multi-party computation, Verifiable Secret Sharing | Secure high frequency smart meter data that can be used for user profiling | Privacy peers are weaker point in the system. If they are compromised. The data is leaked |
| 8 | A New Approach to Secure Aggregation of Private Data in Wireless Sensor Networks [9] | Shared Secret between sink and aggregator | additive functionality of complex numbers, Share secret between sink and source node for data masking | Involve repeated encryption and decryption at each node for each received message |
| 9 | An Efficient Secure Data Aggregation Scheme for Wireless Sensor Networks [17] | Elliptic Curve Okamoto-Uchiyama (EC-OU), Elliptic Curve Digital Signature Algorithm (ECDSA) | Ensure data privacy and integrity that is secure against node compromise attack | Public Key Cryptography for Encryption |
| 10 | Efficient Concealed Data Aggregation with Verification in Wireless Sensor Networks [18] | Elliptic Curve El Gamal (ECEG) | Verify data at each hop and thus provide early detection of attacks. Use homomorphic feature of ECEG | ECEG is power hungry cryptography |
| 11 | PPM-HDA: Privacy-Preserving and Multifunctional Health Data Aggregation With Fault Tolerance [20] | Boneh-Goh Nissim Cryptosystem, Hashing | Data is Encrypted and hashed and then sent to the Social Spot | Boneh Goh-Nissim Cryptosystem is Asymmetric Cryptosystem which is computationally expensive |
| 12 | A Secure Privacy-Preserving Data Aggregation Model in Wearable Wireless Sensor Networks [21] | Public Key Cryptosystem | Use three sets of keys. BS-Cluster Head, CH-Nodes, Nodes- nodes | Key distribution is difficult and costly cryptography |

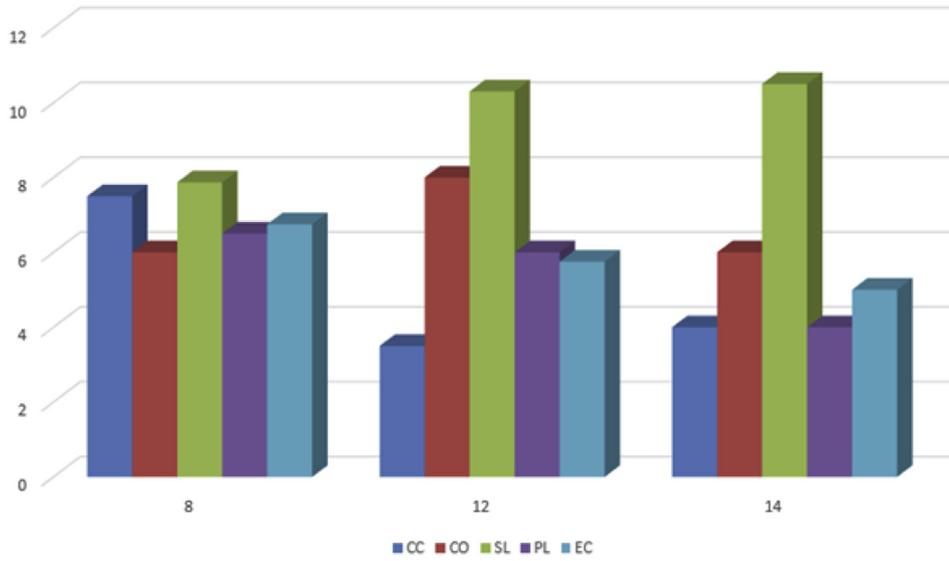

Fig. 3. Comparison of schemes in [8,11,13].

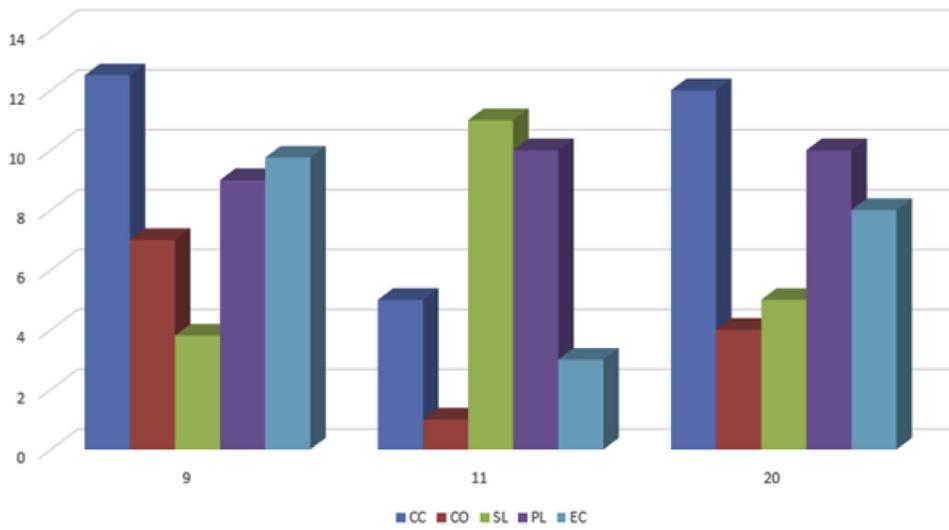

Fig. 4. Comparison of schemes in [5,10,19].

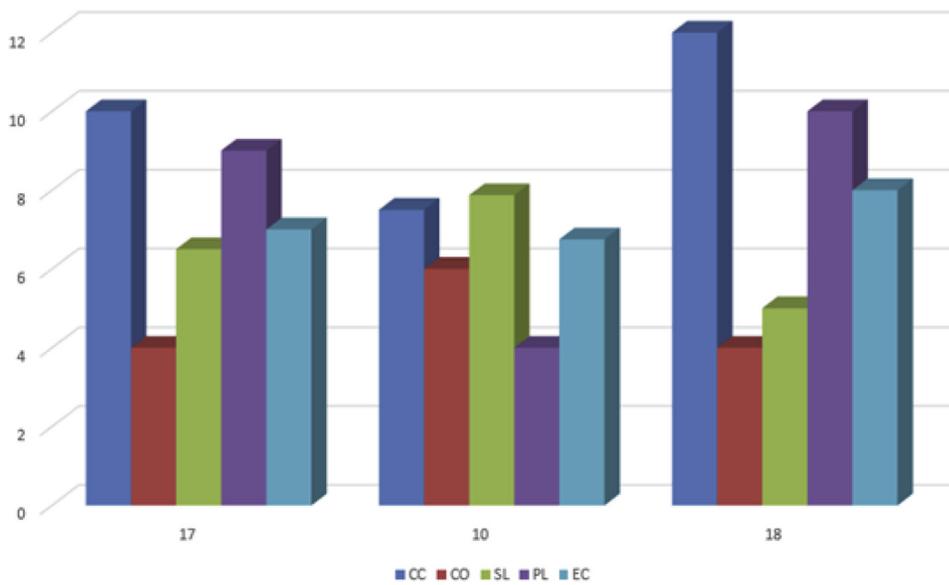

Fig. 5. Comparison of schemes in [9,16,17].

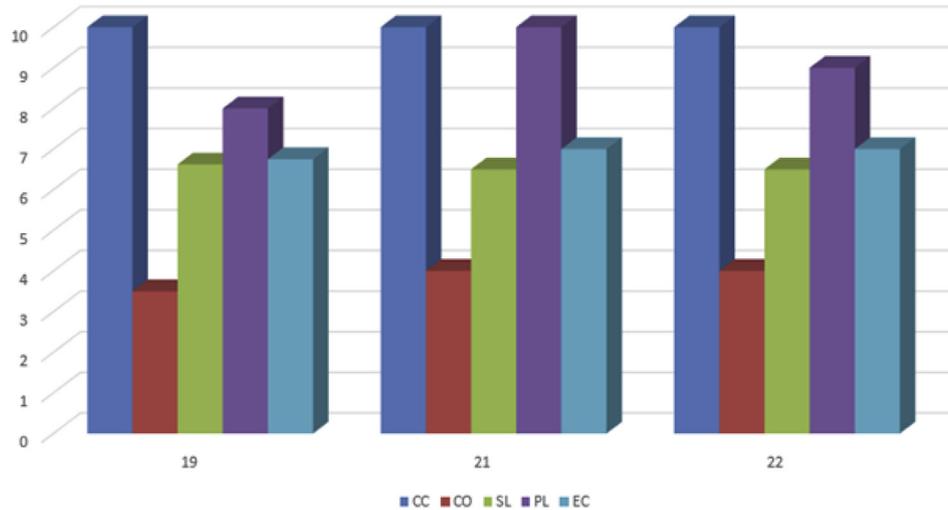

Fig. 6. Comparison of schemes in [18,20,21].

The remaining techniques [16–20] do not produce communication overhead as they neither use data slicing or secret sharing nor it generate extra traffic in key distribution. Beside these Zhang et al. [21] gives a bit high communication overhead as it distributes too many keys-i.e. separate Key pair for Node–Node, aggregator–aggregator and aggregator to BS communication.

### 3.3. Privacy level and privacy against malicious aggregator

Hayouni et al. [5] provide a high level of privacy and also secure against malicious aggregator. Yip et al. [10] do not provide Privacy against Aggregator (PA) while [13], [11], and [8] provide PA. PL of [10] is high due to the preimage resistant property of IHF while [13] just subtract a vector from original data to ensure privacy. PL of both [11] and [8] is medium because data can be leaked if a node successfully get all the shares. Bista et al. [9] provide high PL but as the data is decrypted by the aggregator and then aggregated it means that it does not provide PA.

The work done in Refs. [16–21] all give a very high level of privacy owing to the use of secure public key cryptosystems. These proposed techniques are also secure against malicious aggregator however their use can be challenging due to their high computational complexity.

### 3.4. Sensor node life and energy consumption

Since sensor node life (SL) is inversely proportional to CC and CO while EC is directly proportional to CC and CO.

Table 2
Comparative analysis.

| Reference | Computational complexity | Sensor node life | Communication overhead | Privacy level | Privacy against aggregator | Energy consumption |
|---|---|---|---|---|---|---|
| Hayouni et al. [5] | Very High | Small | High | High | Yes | High |
| Yip et al. [10] | Medium | – | Very Small | High | No | Small |
| Finster and Baumgart [11] | Low | High | High | Medium | Yes | Medium |
| Othman et al. [19] | Very High | Very Small | Small | Very High | Yes | Very High |
| Kumar and Madria [8] | Medium | Medium | Medium | Medium | Yes | Medium |
| Zhang et al. [16] | High | Low | Small | High | Yes | High |
| Callegari et al. [13] | Low | – | Medium | Medium | Yes | Small |
| Bista et al. [9] | Medium | Small | Medium | High | No | Medium |
| Othman et al. [17] | Very High | Small | Small | High | Yes | High |
| Rafik and Mohammed [18] | High | Small | Small | High | Yes | High |
| Han et al. [20] | High | Small | Small | High | Yes | High |
| Zhang et al. [21] | High | Small | Small | High | Yes | High |

Therefore Hayouni et al. [5] have small SL and high energy consumption (EC). Finster and Baumgart [11] high SL and medium EC. Kumar and Madria [8] have both medium SL and EC. Yip et al. [10] and callegary et al. [13] can both have high sensor life. Due to repeated encryption and decryption of message Bista et al. [9] provide small SL and medium EC.

The algorithm proposed in Refs. [16,18,21] use PKC for encryption only and thus these algorithms have small sensor life and high energy consumption. Moreover, the algorithms proposed in Refs. [17,19,20] use PKC both for encryption and signature and thus due to their very high computational cost SL is very small and consume a lot of energy (Table 2).

## 4. Conclusion

Privacy-preserving data aggregation (PPDA) has gained the attention of researchers from the last decade. With the notion of the Internet of Things (IoT), it is gaining more devotion due to its wide application area. In this paper, we surveyed recent contributions of researchers in PPDA in resource constrained IoT sensor nodes. We provide a comparative analysis of these state of the art techniques based on 6 performance parameters- i.e. computational complexity, communication overhead, sensor node life, privacy level, privacy against malicious aggregator and energy consumption by each sensor node. We concluded that Asymmetric cryptosystem being very secure, still not a viable solution for resource constrained nodes as these types of cryptosystem to have very high computational complexity. Data slicing and secret sharing are one of the promising mechanism for PPDA but it causes communication overhead and thus shrinks sensor node life.

This contribution will help the researchers to study and understand the recent work in PPDA and will guide them to where they should spend their efforts to propose more efficient lightweight privacy-preserving techniques for the resource-constrained Internet of things sensor nodes.

## Acknowledgment

This work is a part of our project "Privacy preserving data aggregation techniques in e-health care system" that is funded by my host institution as a part of master degree thesis.

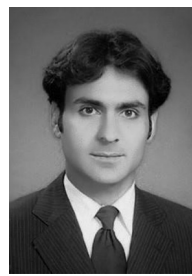

**Inayat Ali** received his Bachelor degree in Telecommunication and Computer Network from COMSATS Institute of Information Technology Abbottabad, Pakistan in March 2015. He is currently the student of Master in Computer Science in COMSATS Institute of Information Technology Abbottabad, Pakistan. His research interest includes Internet of Things, Security, Software Defined Networking, Communication protocols engineering and Mobile Ad hoc Networks.